# Interference fringes in a nonlinear Michelson interferometer based on spontaneous parametric down-conversion

CHEN YANG,[1,2] ZHI-YUAN ZHOU,[1,2,5] LIU-LONG WANG,[3] YAN LI,[1,2] SHI-KAI LIU,[1,2] ZHENG GE,[1,2] XIAO-CHUN ZHANG,[4] QING TANG,[4,6] GUANG-CAN GUO,[1,2] AND BAO-SEN SHI [1,2,7]

[1] *CAS Key Laboratory of Quantum Information, University of Science and Technology of China, Hefei, Anhui 230026, China*
[2] *Synergetic Innovation Center of Quantum Information & Quantum Physics, University of Science and Technology of China, Hefei, Anhui 230026, China*
[3] *Department of Applied Physics, School of Physics and Materials Science, Anhui University, Hefei 230039, China*
[4] *ENNOVA Institute of Life Science and Technology, ENN Group, Langfang 065001, China*
[5] *zyzhouphy@ustc.edu.cn*
[6] *tangqing@enn.cn*
[7] *drshi@ustc.edu.cn*

**Abstract:** Quantum nonlinear interferometers (QNIs) can measure the infrared physical quantities of a sample by detecting visible photons. A QNI with Michelson geometry based on the spontaneous parametric down-conversion in a second-order nonlinear crystal is studied systematically. A simplified theoretical model of the QNI is presented. The interference visibility, coherence length, equal-inclination interference, and equal-thickness interference for the QNI are demonstrated theoretically and experimentally. As an application example of the QNI, the refractive index and the angle between two surfaces of a BBO crystal are measured using equal-inclination interference and equal-thickness interference.

## 1. Introduction

Interferometers have been the most useful and important metrology tools. A novel type of interferometer known as a nonlinear interferometer [1-5] was recently realized. Different from the traditional interferometers, the interference in the novel interferometers is caused by the indistinguishability between photons from two nonlinear sources. The nonlinear sources can be realized in many physical systems based on four-wave mixing or parametric down-conversion, including atomic ensemble [6], integrated silicon photonic chip [7], nonlinear optical fiber [8, 9] and bulk χ(2)-nonlinear medium [10-12].

In this work, we focus on the nonlinear interferometers based on spontaneous parametric down-conversion (SPDC) [13, 14] of bulk χ(2)-nonlinear crystals. Because SPDC is usually used in a quantum source [15], we refer to these interferometers as quantum nonlinear interferometers (QNIs) to distinguish them from other nonlinear interferometers (using other nonlinear effects, for example, the parametric amplification). In recent years, these interferometers have seen growing interest due to the properties of correlations and nonlinearity. Owing to the correlations of momentum and frequency, information can be extracted by a photon never detected [10, 11]. On the other hand, owing to the nonlinearity, two photons in the two arms of a QNI can be frequency highly nondegenerate. These allow that an infrared photon in the interferometer is used to contact samples and extract information, while another visible photon is detected. Because the performances of infrared light detectors are generally inferior to that of visible light detectors, an advantage of QNIs is that one can take advantage of the high quality of visible light detectors to record infrared information. Another advantage is that one can easily tune the photon wavelength in a relatively wide band by changing the phase-matching conditions in some nonlinear sources. For example, silicon chips and some biological tissues are high transmittances in infrared bands, so QNIs can be used to extract their image information by infrared photons and recorded by visible photons using silicon intensified CCD (ICCD) cameras [16]. Moreover, these interferometers also have applications in imaging [10, 16-20], metrology [6, 12], polarimetry [21, 22], optical coherence tomography [23-25], spectrum shaping [26], spectroscopy [11, 18], testing the complementarity principle [27, 28], and measuring bi-photon correlations [29, 30].

References [2] and [30] reported equal-inclination interference in their QNIs and proved that the changing of interference pattern is characterized by a combination of wavelengths of signal and idler photons. In this work, we use a Michelson geometry of QNI [12, 16, 18-20, 23] that is different from the configuration in Ref. [2] and [30], to study both equal-inclination and equal-thickness interference theoretically and demonstrate the theoretical model experimentally. In our configuration, a smaller equivalent wavelength for the equal-inclination interference is obtained compared to Ref. [2]. Moreover, we present two applications of metrology for equal-inclination interference and equal-thickness interference. Our work provides a method to measure the optical properties of crystals. Especially in the mid-infrared band, traditional Michelson interferometers may not applicable in some laboratories for lack of mid-infrared sources or detectors. In these cases, our scheme could be a good alternative.

## 2. Theoretical model

### 2.1 The basic phenomenon and principle of QNI

Our experiment is based on the SPDC of a second-order nonlinear crystal. A simplified schematic shown in Fig. 1 (a) is used to introduce the basic phenomenon and principle of QNI. A monochromatic pump beam sequentially passes through two identical nonlinear crystals, and the down-converted idler and signal photons generated by the two crystals are aligned. The interference is caused by the indistinguishability: one cannot say which crystal the photon pairs are generated from. Although the idler photons also have an interference effect in this scheme, we only focus on the detection probability of signal photons and ignore the idler photons, because in our experiment we refer to the idler photons as infrared photons that cannot be detected by a standard silicon CCD.

This QNI has only one output, unlike the traditional interferometers and the nonlinear interferometers reported in [2, 30, 31], where the energy of light redistributes between two outputs. Here, the energy redistributes nonlinearly between pump photons and the down-converted photons, i.e., the photon number of the down-converted photons is not conserved in the interferometer. The two spatial separated crystals should be treated as a whole system and the total average number of photon pairs generated in the whole system has the form $2|\varepsilon|^2 (1+\cos\phi)$, where $|\varepsilon|^2$ is the probability of photon pair generated from one crystal; the phase $\phi$ is determined by the length and refractive index of the medium between the two crystals.

The principle of the QNI in Fig. 1 (a) is given analytically in the supplementary materials. Here, the interference can be understood using a simple physical picture: the two crystals can be regarded as one long crystal with double length and a phase jump $\phi = k_p l_p - k_s l_s - k_i l_i$ in the middle, where monochromatic plane waves are considered, $k_j = n_j \omega_j / c$ represents the wave vector, where the subscripts $j = $ p,s,i indicate pump, signal, and idler; $\omega$, $c$, and $n$ represent angular frequency, speed of light, and refractive index of air; $l$ represents the path length between the two crystals. The phase jump can be inserted in the phase-matching condition [32] because they have similar form when the three photons in SPDC process have equal path length $l$. Assuming that the phase-matching condition is satisfied, $4|\varepsilon|^2$ photon pairs can be generated when $\phi$ is equal to zero (the efficiency of SPDC is approximately proportional to the square of the crystal length [14, 33]). However, when $\phi$ is equal to $\pi$, it is equivalent to the case of phase mismatch, therefore, no photon pairs are generated. In this sense, this setup can also enable the nonlinear interference of three-wave mixing process [34].

The output bi-photon state from the QNI is given by

$$|\psi_{out}\rangle \approx |0\rangle + \varepsilon \exp(ik_p l_p) \int d\vec{k}_s d\vec{k}_i C(\vec{k}_s, \vec{k}_i) \left[1 + e^{-i\phi(\vec{k}_s, \vec{k}_i)}\right] |1_{k_s} 1_{k_i}\rangle \quad (1)$$

Here, $\varepsilon$ is a constant and the modulus square $|\varepsilon|^2$ represents the total probability of photon pair generation, while $|\varepsilon C(\vec{k}_s, \vec{k}_i)|^2$ represents the probability of photon pair generation where the signal and idler photons have wave vectors of $\vec{k}_s$ and $\vec{k}_i$ respectively; $C$ is a conditional probability of signal and idler photons to have wave vectors of $\vec{k}_s$ and $\vec{k}_i$ given the photon pair has been generated; $C$ satisfies $\int d\vec{k}_s d\vec{k}_i |C(\vec{k}_s, \vec{k}_i)|^2 = 1$. The function $C(\vec{k}_s, \vec{k}_i)$ can be calculated in the interaction picture by integrating the interaction Hamilton [33]. The distribution of $C(\vec{k}_s, \vec{k}_i)$ is closely related to the phase-matching condition, therefore, $C(\vec{k}_s, \vec{k}_i)$ also determines the momentum correlation properties of the down-converted photons.

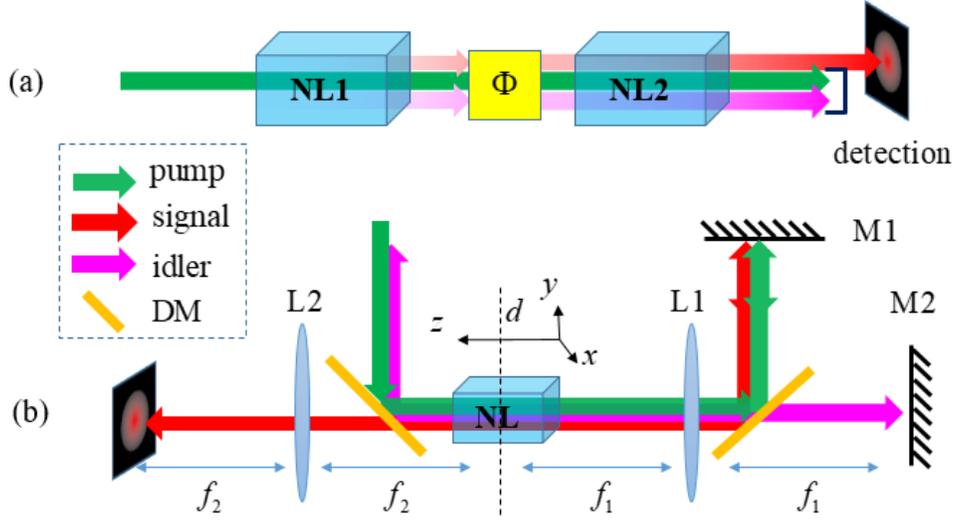

Fig. 1 (a) Schematic of QNI based on second-order nonlinear crystals (NLs). (b) Schematic of the modified QNI with Michelson geometry. Assume z-axis parallel to the direction of light propagation, and set the origin at the front focal plane of L1. $d$ is the position of NL in the z-axis. $\Phi$: transmission medium that induces phase difference among three waves; NL: nonlinear crystal; L: lens; DM: dichroic mirror; M: mirror; $f$: focal length.

### 2.2 QNI with a Michelson geometry

The schematic for the nonlinear interference in Fig. 1 (b) has a modified geometry compared with that in Fig. 1 (a), in which the pump beam and the signal and idler photons are reflected back into the same crystal. This schematic resembles traditional Michelson interferometers, which allows better flexibility because the phase factors can be introduced independently for signal and idler photons as well as for the pump light [1, 3]. Moreover, this scheme has been used for testing the complementarity principle [27, 28]. Our experiment is performed using this scheme, therefore, the following theoretical models are based on this scheme. In this schematic, a lens (L1) is positioned at the center between the crystal and the mirrors (M1 and M2); the crystal and the mirrors located at the front and rear focal plane of L1; L1 and M1 (or M2) form a 4-f imaging system where the photons pass through L1 twice and the object and image planes coincide at the crystal. Another lens (L2) is used as a Fourier translator that maps the spatial frequency components to the spatial point on the detection plane. Therefore, measuring the photon number of each spatial point is equivalent to measuring the photon number of each spatial frequency component, then, considering Eq. (1), we have

$$\langle n(\vec{k}_s) \rangle = \int d\vec{k}_i 2|\varepsilon C(\vec{k}_s, \vec{k}_i)|^2 \left[1 + \cos\phi(\vec{k}_s, \vec{k}_i)\right]$$
$$\approx 2|\varepsilon|^2 \int d\vec{k}_i \left[1 + \cos\phi(\vec{k}_s, \vec{k}_i)\right] \delta^{(3)}(\vec{k}_s, \vec{k}_i) P(\vec{k}_s). \quad (2)$$
$$= 2|\varepsilon|^2 \left[1 + \cos\phi(\vec{k}_s)\right] P(\vec{k}_s)$$

In Eq. (2), the independent variable $\vec{k}_j$ can be also expressed as $(\omega_j, \vec{\kappa}_j)$, where $\vec{\kappa}_j$ is the transverse component of $\vec{k}_j$ ($j = s, i$). The approximation $|C(\vec{k}_s, \vec{k}_i)|^2 \approx \delta^{(3)}(\vec{k}_s, \vec{k}_i) P(\vec{k}_s)$ is based on the ideal energy and transverse momentum entanglement under the conditions of narrow pump linewidth and large pump beam width [30], where the delta function, representing the conditional probability of the idler photon to have wave vector of $\vec{k}_i$ given the signal photon has a wave vector of $\vec{k}_s$, is defined by $\delta^{(3)}(\vec{k}_s, \vec{k}_i) = \delta^{(1)}(\omega_s + \omega_i - \omega_p)\delta^{(2)}(\vec{\kappa}_s + \vec{\kappa}_i)$ including the two entanglement relations ($\omega_s + \omega_i - \omega_p = 0$ and $\vec{\kappa}_s + \vec{\kappa}_i = 0$); here, $\delta^{(1)}$ and $\delta^{(2)}$ are one-dimension and two-dimension Dirac delta function respectively; $P(\vec{k}_s) = \int d\vec{k}_i |C(\vec{k}_s, \vec{k}_i)|^2$, satisfying $\int d\vec{k}_s P(\vec{k}_s) = 1$, is the probability of signal photon to have wave vector $\vec{k}_s$. The phase difference $\phi(\vec{k}_s, \vec{k}_i)$ is dependent on the optical path of both the signal and idler photons. After the integral in the second line over the idler mode $\vec{k}_i$, including $\omega_i$ and $\vec{\kappa}_i$, the phase difference $\phi(\vec{k}_s) = \phi(\omega_s, \vec{\kappa}_s)$ only depends on the signal mode $(\omega_s, \vec{\kappa}_s)$ because of the two entanglement relations.

Eq. (2) is the basis of the following models where the form of phase difference $\phi(\vec{k}_s)$ determines the type of interference. Three forms of $\phi(\vec{k}_s)$ are concerned in this work and they are corresponding to three cases of interference: (1) the two arms are both perfect 4-f system and well aligned; (2) the 4-f system is defocused and equal-inclination interference occurs; (3) the mirror in one of the arms is slightly tilted and equal-thickness interference occurs. In the first case, all the spatial modes have equal optical paths and the phase difference is approximately independent of the spatial modes. In other words, the phase difference is the same as that in the case of single-mode assumption and $\phi(\vec{k}_s) = k_p l_p - k_s l_s - k_i l_i$. In this case, the interference pattern is a uniform bright or dark spot. In section 2.3, the mean photon number on the detection plane and the expression for interference visibility are presented based on case (1); case (2) and (3) are studied in section 2.4 and 2.5. In addition to these three cases, the function $P$ also induce interference fringes when the arm difference is not zero, which is introduced in the supplementary materials.

*2.3 Mean photon number on the detection plane and interference visibility*

Considering that L1 and L2 are used as Fourier translators, we obtain mapping relations between coordinates and transverse wave vectors, $\vec{\rho}_s = \vec{\kappa}_s f_2 / k_s$, and $\vec{\rho}_i = \vec{\kappa}_i f_1 / k_i$, where $\vec{\rho}_s$ is the coordinate on the detection plane and $\vec{\rho}_i$ is the coordinate on M2. Based on the point-to-point mapping relation between $\vec{\rho_s}$ and $\vec{\kappa_s}$, the mean photon number $\langle n_d(\vec{\rho_s}) \rangle$ at the position of $\vec{\rho_s}$ on the detection plane is equal to the mean number of photons with transverse wave vector $\vec{\kappa_s}$, which is given by the integral over signal frequency

$$\langle n_d(\vec{\rho_s}) \rangle = \langle n(\vec{\kappa_s}) \rangle = \int d\omega_s \langle n(\omega_s, \vec{\kappa_s}) \rangle = 2|\varepsilon|^2 \int_0^\infty d\omega_s [1+\cos\phi_0] P(\omega_s, \vec{\kappa_s}), \tag{3}$$

where $\langle n(\omega_s, \vec{\kappa_s}) \rangle$ and $P(\omega_s, \vec{\kappa_s})$ are the same as $\langle n(\vec{k_s}) \rangle$ and $P(\vec{k_s})$ defined in Eq. (2) respectively. Considering the Michelson interferometer shown in Fig. 1(b), we assume that the arm of the signal and pump photons has a length of $l_1$, i.e. $l_s = l_p = l_1$, another arm has a length of $l_2$, i.e. $l_i = l_2$. In this case, the phase difference is given by

$$\phi_0 = 2k_p l_1 - 2k_s l_1 - 2k_i l_2 = k_i \Delta l + \phi_1, \tag{4}$$

where $\Delta l = 2l_1 - 2l_2 + 2l'$ represents the optical path difference (OPD) between the two arms; the constant $l' = \int_0^{l_1} dz [n_s(z) - n_i(z)]/n_i(z)$ and $\phi_1 = 2\int_0^{l_1} dz [n_p(z) - n_s(z)] \omega_p / c$ are the compensation for the additional OPD caused by the dispersion of air and optical elements; $n_j(z)$ represents the refractive index of air or optical elements between the crystal and the mirror M1 or M2.

The spectral distribution of photons has the form [35]

$$P(\omega_s, \vec{\kappa_s}) = \frac{1}{\Delta\omega} \text{sinc}^2 \left[ 2\pi \frac{\omega_s - \omega_{s0}(\kappa_s)}{\Delta\omega} \right] = \frac{1}{\Delta\omega} \text{sinc}^2 \left[ 2\pi \frac{\omega_{i0}(\kappa_s) - \omega_i}{\Delta\omega} \right], \tag{5}$$

where $\text{sinc}(x) = \sin x / x$; $\omega_{s0}(\kappa_s)$ and $\omega_{i0}(\kappa_s) = \omega_p - \omega_{s0}(\kappa_s)$ are the center angular frequency of signal and idler photons, respectively, depending on the transverse wave vector $\kappa_s = |\vec{\kappa_s}|$; they satisfy an approximate parabolic relation $\omega_{s0}(\kappa_s) - \omega_{s0} = \omega_{i0} - \omega_{i0}(\kappa_s) = b\kappa_s^2$, where $b$ is a constant factor and is determined by the dispersion properties of the crystal (see Supplementary 1); $\omega_{s0} = \omega_{s0}(0)$ and $\omega_{i0} = \omega_{i0}(0)$ are the center frequencies in the collinear phase-matching case; $\Delta\omega$, denoting frequency spectral full width of both signal and idler photons and satisfying $P(\omega_{s0} \pm \Delta\omega/2, \vec{\kappa_s}) = 0$, is determined by the length and dispersion properties of the nonlinear crystal; moreover, $\Delta\omega$ is also slightly dependent on the transverse vector $\vec{\kappa_s}$, which can be calculated based on the non-collinear phase-matching condition; Fig. S2 (a) in the supplementary materials shows the function $P$ in our experiment. Because an aperture or narrow band filter is usually used in an experiment, $\Delta\omega$ can be approximately treated as a constant that is calculated in the collinear phase-matching case. After the integral calculation in Eq. (3), where the integral variable is transformed into $\omega_i$ using the relation $\omega_s + \omega_i = \omega_p$, we have

$$\langle n_d(\vec{\rho_s}) \rangle = 2|\varepsilon|^2 \left[ 1 + \text{tri}\left(\frac{\Delta\omega \Delta l}{\pi c}\right) \cos\left(\frac{\omega_{i0}}{c} \Delta l - \phi_2(\rho_s, \Delta l) + \phi_1 \right) \right], \tag{6}$$

where the phase $\phi_2 = \frac{b\omega_{s0}^2}{f_2^2 c^3} \rho_s^2 \Delta l$ can induce ring-like fringes, which is referred as angular-spectrum-dependent (ASD) interference and is introduced in Supplement 1; the triangle function is defined as

$$\mathrm{tri}(x) = \begin{cases} 1-|x|, & -1 < x < 1 \\ 0, & \text{otherwise} \end{cases}. \tag{7}$$

Then, the visibility is given by

$$V = \frac{\langle n_s \rangle_{\max} - \langle n_s \rangle_{\min}}{\langle n_s \rangle_{\max} + \langle n_s \rangle_{\min}} \approx \mathrm{tri}\left(\frac{\Delta l \Delta \omega}{4\pi c}\right) \approx \mathrm{tri}\left(\frac{\Delta l}{2\lambda_i^2 / \Delta \lambda_i}\right), \tag{8}$$

where $\Delta \lambda$ represents the full linewidth of photons. When the OPD is equal to $\pm 4\pi c / \Delta \omega$ the visibility becomes zero, therefore, the coherence length of both the signal and idler photons in the QNI is $8\pi c / \Delta \omega$, which is equivalent to

$$l^c = 4\lambda_i^2 / \Delta \lambda_i = 4\lambda_s^2 / \Delta \lambda_s. \tag{9}$$

To observe interference phenomena, the coherence length of photons should be longer than the OPD, i.e. $l^c \gg |l_s - l_i|$; the coherence length is usually short enough, therefore, we assume that the arm difference $\Delta l$ cannot cause defocusing of the 4-f system. Other condition for this interference should be noted is that the coherence length of pump light should be far longer than the distance between the two crystals, i.e. $l_p^c \gg l_p$. The complex factor $\exp(ik_p l_p)$ in Eq. (1) depends on pump wavelength, therefore, the integration of pump wavelength reduces the interference visibility when the linewidth of the pump light is not narrow enough. In this work, we assume that the pump light is an ideal monochromatic plane wave.

### 2.4 Equal-inclination interference

If the 4-f system is defocused, circular fringes of equal-inclination interference occur. In the case that the optical path of idler arm adds a distance $d$, the phase difference in Eq. (2) is given by $\phi \approx 2\pi d \theta_i^2 / \lambda_i + \phi_0$ [2], where $\theta_i \approx \kappa_i / k_i = \kappa_s \lambda_i / k_s \lambda_s$ is the incident angle of idler photons (angle between the wave vector and the light propagation direction). The first term is the equal-inclination term, which reflects the transverse phase difference and creates the ring-like interference fringes. The second term is the longitudinal phase difference, which is determined by Eq. (4) and does not create interference fringes. A similar equation can be obtained if the signal arm changes.

In our QNI scheme, the photons emit from the crystal in the form of plane waves (the angular spectrum of the plane waves is shown in Fig. S2 in the supplementary materials), and the lens L1 is used as a Fourier translator; if a plane wave with a transverse wave vector $\vec{\kappa_i}$ is incident on L1, it gets converted into a spherical wave that converges to a point $\vec{\rho_i}(\vec{\kappa_i})$ on the mirror M2. Therefore, the equal-inclination interference fringes are only sensitive to the distance between the crystal and L1. The OPD $\Delta L$ included in $\phi_0$ makes almost no contribution to the equal-inclination interference term. In the experiment, we move the crystal a distance $d$ along the beam propagation axis to obtain equal-inclination interference fringes. Under the approximation that the divergence angle of the pump beam is ignored, the introduced phase shift can be expressed as

$$\phi \approx 2\pi d \left( \frac{\theta_i^2}{\lambda_i} + \frac{\theta_s^2}{\lambda_s} \right) + \phi_0 \approx 2\pi d \frac{\theta_s^2}{\lambda_{eq}} + \phi_0, \tag{10}$$

where $\lambda_{eq} = \frac{\lambda_s \lambda_p}{\lambda_i}$ is defined as the equivalent wavelength. The relation $\theta_s / \theta_i = \lambda_s / \lambda_i$ is used in the calculation. The relation $\phi = 2N\pi$ $(N = 0, 1, 2...)$ determines the constructive interference order. The transverse positions of bright rings vary with the phase difference $\phi_0$ that is given by Eq. (4).

*2.5 Equal-thickness interference*

When the crystal is located at the position of $d=0$, i.e. no equal-inclination interference fringes appear, equal-thickness interference occurs if one of the mirrors is slightly tilted, or in an equivalent case that a gradient thickness plate is inserted in one of the two arms. The interference pattern is formed by straight fringes that are perpendicular to the gradient direction. If the thickness varies along the x-axis, the phase difference is given by $\phi = 2nk_i h(x)$, where $h$ represents the thickness of the sample. The generalized expression is given by $\phi = 2nk_i h(\vec{\rho_i})$, where $\vec{\rho_i}$ is the coordinate on the plane of M2 and $n$ is the refractive index of the plate. The gradient thickness plate in front of and close to M2 can be seen as a phase object, and the interference pattern is actually its phase image. This interference effect can be equivalently interpreted using the theory of imaging in Ref. [36]. Given $\theta_i \approx \rho_i / f_1 \approx \kappa_i / k_i$ and $\theta_s \approx \rho_s / f_2$, the transformation relation of the coordinate on the plane of M2 and the detection plane is given by

$$\vec{\rho_i} = \vec{\rho_s} \frac{f_1 \lambda_i}{f_2 \lambda_s}. \tag{11}$$

Then the phase difference becomes

$$\phi = 2(n-1)k_i h\left(\vec{\rho_s} \frac{f_1 \lambda_i}{f_2 \lambda_s}\right). \tag{12}$$

By substituting Eq. (12) into Eq. (2), one can predict the interference patterns. On the other hand, one can perform equal-thickness interference to measure the thickness variation of a plate sample using Eq. (12).

### 3. Experimental setup

A schematic of our experimental setup is shown in Fig. 2. The 525.2-nm light beam of the CW pump laser is generated in a single-pass sum-frequency generation (SFG) process [37] (the SFG source is not shown in Fig. 2). In the SFG source, the wavelengths of the two pump beams are 1540 nm and 797 nm respectively and the linewidths are 10-kHz and 200-kHz respectively. Because both the pump beam have narrow linewidth, the 525.2-nm SHG laser beam also has a narrow linewidth (< 210 kHz). All the three beams are vertically polarized and satisfy type-0 phase-matching condition. The SFG laser beam is collected into a single-mode fiber and exits through a fiber collimator (the FC in Fig. 2). Because the idealized plane-wave pump in the SPDC is a necessary condition of strict transverse momentum correlations, the pump beam for SPDC is collimated by a lens group; its width is of order 400 μm and its power is of order 60 mW that is the maximum power we can obtain. Two PPKTP crystals are used in the experiment, one for SFG and the other for SPDC. The two crystals have the same parameter values: their dimensions are 1 mm × 2 mm × 10 mm, and their grating periods are 9.34 μm. The temperature of the crystal used during SFG is set at 24 ℃, which is an optimum temperature for SFG. The temperature of the crystal used for SPDC is set at 29 ℃. This temperature is determined by performing difference-frequency generation (DFG) between the 525.2-nm and 1540-nm laser beams. The two temperatures are different because the laser focusing conditions are different in the two crystals; the pump beams are tightly focused in the SFG process, while the pump beams are collimated in the DFG or SPDC process; to maximize the SFG efficiency for focused Gaussian beams, a mismatch should be introduced [32]. Because SPDC is the inverse of SFG, the central wavelength of the idler and signal photons are approximately 1540 nm and 797 nm. The pump, signal, and idler photons are split through a long-pass dichroic mirror (DM2), where the idler photons pass through DM2 while the pump and signal photons are reflected. After the second pass of the crystal, the pump photons pass through another short-pass dichroic mirror (DM1), while the idler and signal photons are reflected (DM1 used in our experiment is different from the DM in Fig. 1(b), where we assume the DM transmit signal photons to clearly illustrate the 4-f imaging system). Before detection on the ICCD, the photons pass through an 800-10 nm band-pass filter (the center transmission wavelength is 800 nm and the FWHM bandwidth is 10 nm), therefore, only the signal photons are detected while the idler photons are discarded. To control the arm difference roughly or finely in the experiment, we fix M2 on the displacement platform and M1 on the piezoelectric transducer (PZT).

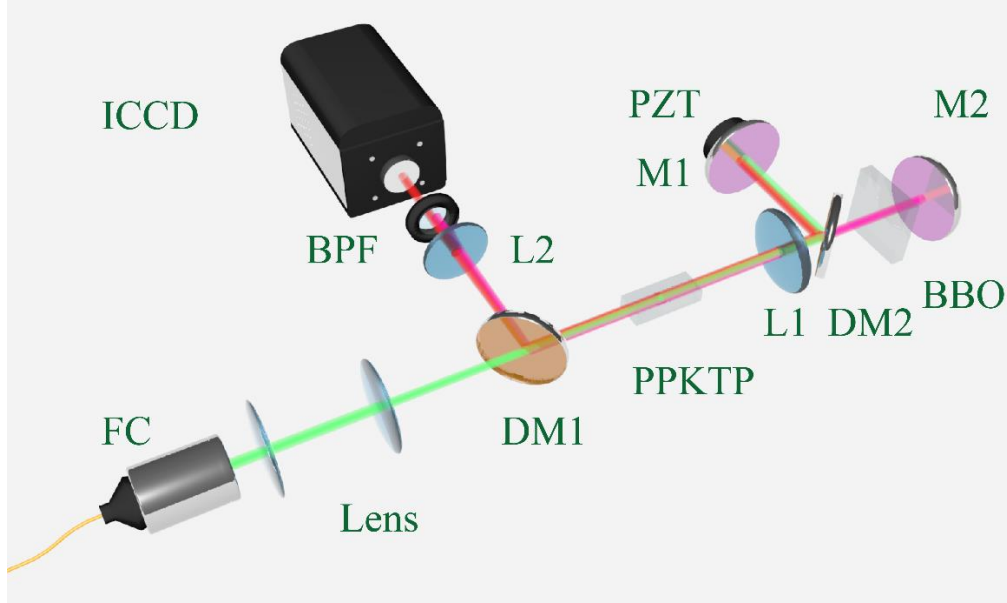

Fig. 2 Experimental setup. The focal lengths of L1 and L2 are $f_1 = 75$ mm and $f_2 = 200$ mm. PPKTP: type-0 periodically poled potassium titanyl phosphate crystal. BBO: Beta-barium borate crystal with dimensions of 5 mm × 5 mm × 0.5 mm. The BBO crystal was inserted when equal-inclination and equal-thickness interference experiments were implemented. BPF: 800 - 10 nm band-pass filter; DM: dichroic mirror; L1, L2: lens; M1, M2: mirrors.

## 4. Results and discussion

In the experiment, we first found the position of M2 where the arm lengths were equal, meanwhile, the coherence length was measured; next, we tested the stability of the interferometer because a relatively stable interferometer is the premise of the reliability of the data; after that, we observed the equal-inclination interference by changing the position of the crystal; then, we studied the influence of OPD on interference in three cases; in the third case, the refractive index of a BBO crystal was measured; finally, an equal-thickness interference was observed and used to measure the angle between the two surfaces of the BBO crystal. As we mentioned above, the equal-thickness interference fringes in our setup are equivalent to the phase imaging of the phase object in the idler arm, therefore, the clear imaging results can prove the reliability of the equal-thickness interference fringes; the imaging results are shown in the supplementary materials.

### 4.1 Coherence length and interference stability

Fig. 3 (a) shows the relation between interference visibility and the position of M2. The inserts are some interference patterns recorded by the ICCD with a 10 s exposure time. To improve the signal-noise ratio and obtain accurate visibilities, we set such a long exposure time, although the exposure time could be much shorter. In order to show the visibility directly, we measure the visibility and record the fringe patterns for equal-inclination interference (corresponding to $d = 6$ in Fig. 4), instead of the interference where the patterns are uniform spots. The background is taken before the data acquisition and is subtracted by the ICCD camera automatically when signals are taken. Nevertheless, given that the lowest stable temperature of our ICCD is -25 °C, there were remain dark counts that reduce the interference visibility. The interference visibility defined by Eq. (8) is measured from the first-order bright ring and dark ring at the center of the pattern. When calculating the visibility, the average dark counts are removed. In order to determine the position of M2 giving an equivalent path, we approximately fit the experimental data with a triangle function based on Eq. (7) and (8). And the origin of horizontal axis has been set at the position with maximum interference visibility obtained by the fitting. The fitting result shows that the maximum interference visibility is $58.8\pm2.7\%$ and the coherence length is $l_{\text{exp}}^{c}=1.28\pm0.06$ mm. The uncertainties were estimated assuming that the counts follow Poisson statistics. The visibility is mainly limited by the losses of the optical elements in the interferometer; the lens L1 is uncoated and has losses of 8.0%, 6.9%, and 7.2% for the pump, signal, and idler photons respectively. To improve the visibility, one can use an off-axis parabolic mirror with a silver coating to reduce the losses. The predicted coherence length is $l_{\text{pre}}^{c}=1.20$ mm from Eq. (9), where the predicted bandwidth used in Eq. (9) is calculated based on quasi-phase-matching condition [32] and the Sellmeier equation we used is from Ref. [38]. The total error comprises the fitting error, the measurement error, and the Sellmeier equation not perfectly matching our crystal.

We then test the interference stability and show the result in Fig. 3 (b). M2 is moved to the position with maximum interference visibility, and the ICCD works in the kinetic mode. In this mode, the ICCD cannot subtract background automatically, so the inserts are vaguer than that in Fig. 3 (a). 4000 interference patterns were continuously recorded and each has 1 s exposure time. We first average the normalized pixels value in the green region shown in the insert, then plot a curve that reflects the light intensity changes with time to assess the stability. The relative root-mean-square intensity fluctuation is 2.4%. This indicates that our interferometer is to some extent stable.

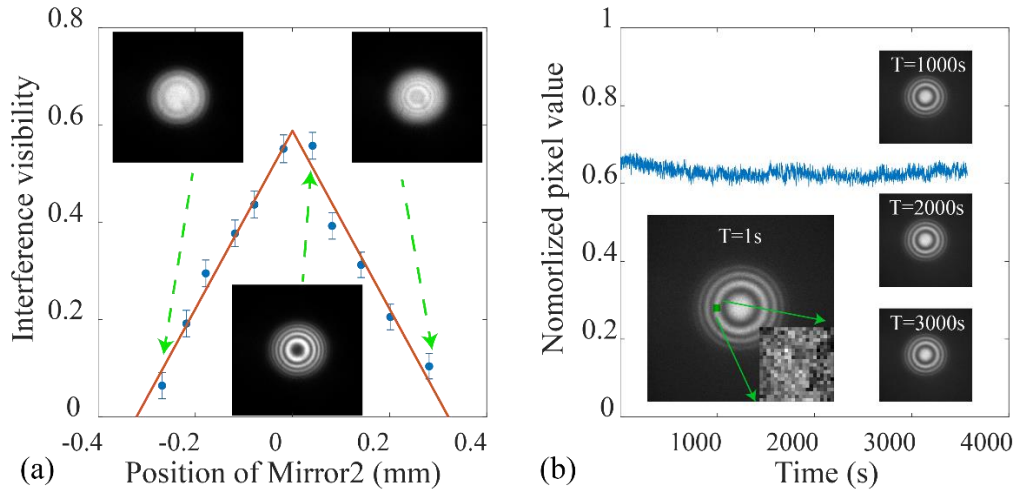

Fig. 3 (a) Interference visibility. The blue spots show the experimental data and errors were estimated assuming Poisson statistics. The data were fitted by a triangle function (red line). The inserts are some interference patterns; their corresponding positions are pointed by the green dashed-line arrows. (b) Interference stability in 4000 s. The values on the blue curve are the average normalized pixel value in the green region shown in the larger insert. The inserts are example interference patterns sampled at 1 s, 1000 s, 2000 s, and 3000 s.

*4.2 Equal-inclination interference*

After the equal arm position is determined, we observe the equal-inclination interference pattern with different crystal positions. The crystal position is set as the origin where the interference pattern is a uniform bright or dark spot. Then, we move the crystal away from L1 and keep the positions of mirrors unchanged. We assume that the coordinate of the crystal (the distance between crystal and origin) is $d$ and the interference patterns with different coordinate $d$ are shown in Fig. 4 (a), where more interference rings appear with increasing coordinate $d$. By substituting Eq. (10) into Eq. (2) and calculating the maximum value of the cosine function, one can obtain the relationship $a\rho_s^2 + \phi_0/2\pi = N$, where $\rho_s = \theta_s f_2$ represents the radius of the ring-like patterns and $N$ is the constructive interference order. The quadratic coefficient given by $a = d/f_2^2 \lambda_{eq}$ determines how fast the relative phase varies with the radius of fringes. The equivalent wavelength can be obtained by fitting the dependence of $a$ to $d$.

Fig. 4 (b) shows experimental data of $(\rho_s, N)$ for different values of $d$. The coefficient $a$ may be evaluated using a second-order polynomial fit to the data. In Fig. 4 (c), the obtained values of $a$ are plotted for different $d$. The red line shows the fitting result from which one obtains the equivalent wavelength $\lambda_{eq} = 254 \pm 11\,\text{nm}$, where the uncertainty was estimated using 95% confidence bounds of the fitting. The experimental value is close to the predicted value of 271.8 nm. The relative error of 6.5 % is caused by the imperfect 4-f system, the dispersion of L1, and the approximation in section 2.4.

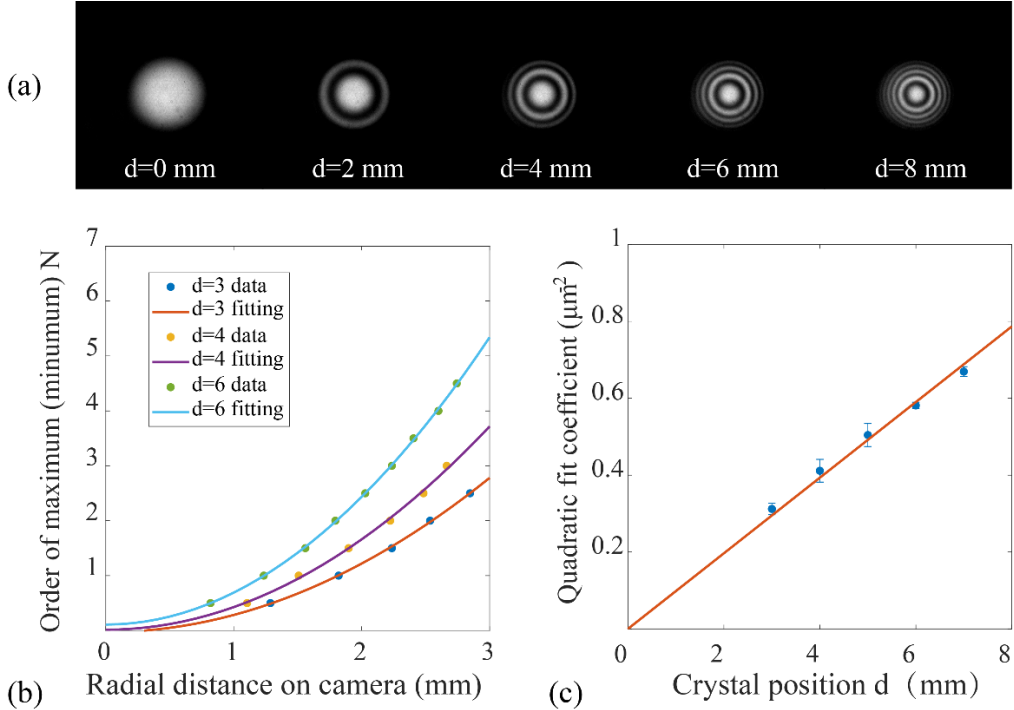

Fig. 4 Results of equal-inclination interference. The lengths of the two arms are equal, while the crystal is shifted away from the mirrors. (a) Interference patterns observed with the crystal positioned at $d = 0, 2, 4, 6, 8\,\text{mm}$. (b) Evaluated radial distance on camera of the minima and maxima for $d = 3, 4, 6\,\text{mm}$. The integer and half-integer orders are corresponding to maxima and minima respectively. (c) The quadratic coefficients $a$ were obtained by fitting the data of $d = 3, 4, 5, 6, 7\,\text{mm}$. Errors were estimated using 95% confidence bounds of the fitting. The data are fitted by a linear function.

The equal-inclination interference could potentially be useful in metrology [2]. In a traditional Michelson interferometer, the variation of fringes can be used to measure the small displacement and changes of refraction index in one of the two arms. In this interferometer, the small displacement and changes of refraction index between the crystal and the lens L1 can be similarly measured by recording the variation of interference fringes. The results shown in Fig. 4 show the dependence of fringes on displacements. Comparing with the measurement in a traditional interferometer, the QNI has advantages as follow: firstly, no reference path is needed (the Michelson geometry is not needed when doing this) and therefore the measurement is not influenced by the errors from reference arm; secondly, the range of measurement is not limited by the coherence length of photons, which is usually very short for photons from SPDC; thirdly, the equivalent wavelength is smaller than the physical wavelengths of the pump photons and is also smaller than the equivalent wavelength in Ref. [2], which means the variation of fringes is more sensitive to the displacement and therefore the measurement sensitivity could be higher than that in traditional interferometers.

*4.3 Three cases for OPD and equal-thickness interference*

In the following, we study the influence of OPD on interference phenomenon, here OPD is determined by arm difference and the crystal is fixed at the position of $d=0$. Three cases for OPD are discussed: (1) the OPD changes slightly, around one or two microns; (2) the OPD changes within the coherence length; (3) the change of OPD can exceed the coherence length. In the first case, the interference patterns are bright or dark spots. We adjust the PZT voltage to record the spot patterns in the cases of different OPDs. Fig. 5 (a) shows that the average photon counts of the spot changes with PZT voltage. We fitted the data using a sine function and the fitting results show that the OPD period is $1527 \pm 8$ nm, where the uncertainty was estimated using 95% confidence bounds of the fitting. From Eq. (4), the theoretical period is equal to the idler wavelength $1540$ nm. The error is from the nonlinearity and inaccuracy of the dependence of thickness to the voltage of the PZT and the measurement error of the voltage. In this case, small phase changes can be measured in practical applications, like in a traditional interferometer.

In the second case, ring-like patterns appear and the number of fringes increases with increasing OPD. The ring-like patterns are similar to the equal-inclination interference patterns, however, the patterns are not created by equal-inclination interference; they are caused by the phase $\phi_2$ in Eq. (6). This phenomenon referred as ASD interference is explained briefly in section 2 in Supplementary 1. In Fig. 3 (a), the number of fringes changes with OPD, which is caused by the ASD interference. The ASD interference has much shorter equivalent wavelength than the equal-inclination interference above, which indicates that the ASD interference has potential in improving measurement sensitivity when the OPD is measured by counting fringes.

In the third case, the coherence length is used as a ruler to measure OPD. For an application example, we inserted a BBO crystal with a length of $h = 0.5$ mm in the idler arm to measure its slow axis refractive index. Fig. 5 (b) shows that the interference visibility curve before and after insertion of the BBO crystal. The red arrow shows the translation direction. The center of the triangle function moves a distance of $d_B = 342 \pm 11\,\mu m$, where the uncertainty was estimated using 95% confidence bounds of the fitting. The refractive index $n_o = 1.683 \pm 0.054$ is obtained by calculating the equation $2(n_o - 1)h = 2d_B$. The predicted refractive index 1.647 is calculated using the Sellmeier equation in ref [39]. The error is also caused by the inaccuracy of the thickness of BBO crystal and the BBO surface being not perfectly vertical to the light propagation direction.

When the BBO crystal is inserted and the PPKTP crystal is fixed at the origin ($d = 0$), the ring patterns become straight fringes (not spot). The interference patterns, shown in Fig. 5 (c), are created by equal-thickness interference because the two surfaces of the crystal are not perfectly parallel. To ensure the accuracy of Eq. (11) and (12), the BBO crystal was very close to the mirror M2 and was about 5mm from it; the distance cannot be shorter, which was prevented by the mounting frame of the crystal (assuming the Gaussian waist for idler photons is 400 μm in the crystal, one can calculate out the waist on the mirror M2 that is of order 92 μm, which indicates that the Rayleigh length is of order 17 mm; therefore, the distance 5 mm and the slight focal shift caused by the BBO crystal is small enough to satisfy the phase imaging approximation in section 2.5). The brightness of fringes reflects the relative phase difference $\phi(\vec{\rho}_s)$ in Eq. (12). Based on Eq. (11) and (12), one can obtain the thickness function

$$h(\vec{\rho}_i) = \frac{\phi(\vec{\rho}_s)+\phi_0}{2(n_o-1)k_i} . \quad (13)$$

The angle of 1.3'±0.1' between the two surfaces is obtained by dividing the thickness difference $\pi/2(n_o-1)k_i$ by the distance between bright and dark strips, where the distance was measured manually on the camera and the uncertainty was derived from the variance of the measurement for the distance. Here, we only demonstrate the measurement practicability, not accuracy, because no standard value of the angle can be used as a reference.

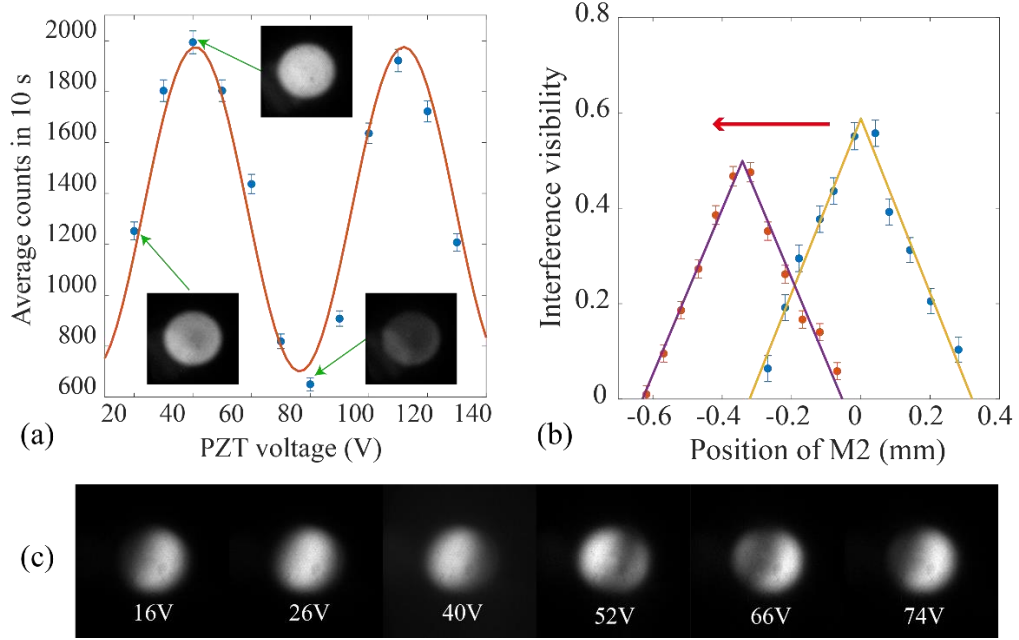

Fig. 5 (a) Interference patterns observed and average counts evaluated by fine translation of the mirror M1 using a PZT (the PZT stretch 11 nm per volt). The blue spots are experimental data and are fitted using a sine function. (b) Interference visibility of equal-inclination interference before and after insertion of the BBO crystal. Errors were estimated assuming Poisson statistics. (c) Equal-thickness interference patterns after insertion of the BBO crystal.

## 5. Conclusion

In summary, a QNI with Michelson geometry is realized. The interference visibility, coherence length, equal-inclination interference, and equal-thickness interference are demonstrated theoretically and experimentally and the experimental results agree with those of our theoretical model. In stressing the study of interference pattern observed by moving crystal, the variation of the common path of the idler and signal photons can prompt equal-inclination interference and the interference fringes vary faster with the increasing common path than those for traditional equal-inclination interference with increasing path difference; this phenomenon is parameterized by an equivalent wavelength and the expression for the equivalent wavelength is presented, which is determined by the wavelengths of all three photons in the SPDC process (or other two photons, because only two wavelengths are independent). The potential application and advantages in metrology is discussed in section 4.2. To improve the parameters of the present interferometer, one can improve the pump power to reduce the exposure time and reduce the loss of elements to improve the interference visibility.

By comparing with the traditional interferometers, the main advantage of QNIs is that they can measure the infrared physical quantities through the detection of visible photons. In this work, we demonstrate that the infrared refractive index of a crystal can be measured using the QNI. Also, the surface irregularities can be measured through equal-thickness interference, like in the traditional interferometers. When the sample is an IR material that has high transmittances in infrared band and low transmittances in visible band, the nonlinear interferometer based on spontaneous down-conversion can provide infrared photons from near-IR to far-IR [40] and take advantage of visible detectors to record the interference fringes. For example, this type of interferometer can be used in the application of microscopes for silicon chips [19]. We believe that the above research for the quantum nonlinear Michelson interferometer and its advantage demonstrated here may not only enrich the knowledge regarding optical interference but also be helpful for the possible applications of infrared metrology and imaging.

**Funding.** National Natural Science Foundation of China (NSFC) (61605194, 11934013, 61525504); Anhui Initiative In Quantum Information Technologies (AHY020200); China Postdoctoral Science Foundation (2017M622003, 2018M642517).

**Disclosures.** The authors declare no conflicts of interest.

**Supplemental document.** See Supplement 1 for supporting content.

**Data availability.** Data underlying the results presented in this paper are not publicly available at this time but may be obtained from the authors upon reasonable request.

# Supplementary Materials

## 1. Principle of the nonlinear Michelson interference based on SPDC

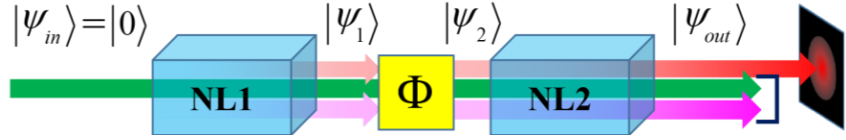

Fig. S1 Schematic of the nonlinear interference based on second-order nonlinear crystals (NLs). See Eq. (S1) ~ (S3) for the quantum states at their corresponding position.

In order to explain the nonlinear interference phenomenon, we first consider the single-mode case for simplification, in which we assume that the SPDC photons are plane waves. The evolution operator of the first nonlinear crystal can be approximately represented by $U_1 = 1 + \varepsilon a_s^+ a_i^+$. Here, the pump field is approximately treated as a classical field and its information is contained in the coefficient $\varepsilon$ that is also proportional to the nonlinear susceptibility. $a_s^+$ and $a_i^+$ are the generation operators of signal and idler photons respectively. Here, the SPDC photon state generated by the first nonlinear crystal can be represented by

$$|\psi_1\rangle = U_1|0\rangle = |0\rangle + \varepsilon|1_s 1_i\rangle. \tag{S1}$$

$|\varepsilon|^2 \ll 1$ is the probability of photon pair generation in the crystal. After propagation between the two crystals, the quantum state becomes

$$|\psi_2\rangle = |0\rangle + \varepsilon \exp(ik_s l_s + ik_i l_i)|1_s 1_i\rangle. \tag{S2}$$

In the second crystal, the evolution operator becomes $U_2 = 1 + \varepsilon \exp(ik_p l_p) a_s^+ a_i^+$. Here, $k_j = n_j \omega_j / c$ represents wave vector, where the subscripts $j = p, s, i$ indicate pump, signal and idler; $\omega$ and $c$ represent angular frequency and speed of light; $n$ and $l$ represent the refractive index of the medium and path length respectively between the two crystals. The final state is given by

$$|\psi_{out}\rangle = U_2|\psi_2\rangle \approx |0\rangle + \varepsilon e^{ik_p l_p}\left(1 + e^{-i\phi}\right)|1_s 1_i\rangle \tag{S3}$$

where the phase difference is expressed as

$$\phi = k_p l_p - k_s l_s - k_i l_i, \tag{S4}$$

In Eq. (S3), we only keep the first-order term for approximation. After the output, the idler photons and pump beam are discarded and only the signal photons are detected. The average signal photon number is given by

$$\langle n_s \rangle = \langle \psi_{out}|n_s|\psi_{out}\rangle = 2|\varepsilon|^2(1 + \cos\phi), \tag{S5}$$

where $n_s$ is the photon number operator of signal photons.

We then generalize the above calculations to the practical case where the multimode photons are considered. In this case, the evolution operator in the crystal can be represented as

$$U = 1 + \varepsilon \int d\vec{k}_s d\vec{k}_i C(\vec{k}_s, \vec{k}_i) a_{\vec{k}_s}^+ a_{\vec{k}_i}^+. \tag{S6}$$

Here, $\varepsilon$ is a constant and the modulus square $|\varepsilon|^2$ represents the total probability of photon pair generation, while $|\varepsilon C(\vec{k}_s, \vec{k}_i)|^2$ represents the probability of photon pair generation where the signal and idler photons have wave vectors of $\vec{k}_s$ and $\vec{k}_i$ respectively; $C$ is a conditional probability of signal and idler photons to have wave vectors of $\vec{k}_s$ and $\vec{k}_i$ given the photon pair has been generated; $C$ satisfies $\int d\vec{k}_s d\vec{k}_i \left| C(\vec{k}_s, \vec{k}_i) \right|^2 = 1$. The function $C(\vec{k}_s, \vec{k}_i)$ can be calculated in the interaction picture by integrating the interaction Hamilton [1]. The distribution of $C(\vec{k}_s, \vec{k}_i)$ is closely related to the phase-matching condition, therefore, $C(\vec{k}_s, \vec{k}_i)$ also determines the momentum correlation properties of the down-converted photons. After some similar calculations, the output state becomes

$$\left| \psi_{out} \right\rangle \approx \left| 0 \right\rangle + \varepsilon \exp(ik_p l_p) \int d\vec{k}_s d\vec{k}_i C(\vec{k}_s, \vec{k}_i) \left[ 1 + e^{-i\phi(\vec{k}_s, \vec{k}_i)} \right] \left| 1_{k_s} 1_{k_i} \right\rangle \tag{S7}$$

## 2. Imaging results in our nonlinear interferometer

In this section, we present additional results from the imaging experiment in our quantum nonlinear interferometer (QNI). The object is a transmissive USAF-1951 glass slide resolution target and is inserted in front of the mirror M2 instead of the BBO crystal (the glass slide is against the mirror and the arm difference should be adjusted when the sample is replaced because of the short coherence length of photons). The results are shown in Fig. S2; (a) and (c) show the constructive images; (b) and (d) show the destructive images; the image in (e) is the difference image of (b) and (d); (f) shows an interference image without band-pass filter used in in the experiment, where the constructive and destructive interferences occur periodically along the radial direction (this phenomenon is explained in the second section). Fig. S2 (a) and (b) show the clear images of the larger test patterns and clear images indicate that the imaging system is not defocused and the phase images of the BBO crystal shown in Fig .5 are reliable. Fig. S2 (c)~(e) show the images of the smaller test patterns. From the lines at the bottom right of Fig. S3 (e), is estimated as 250 micrometers.

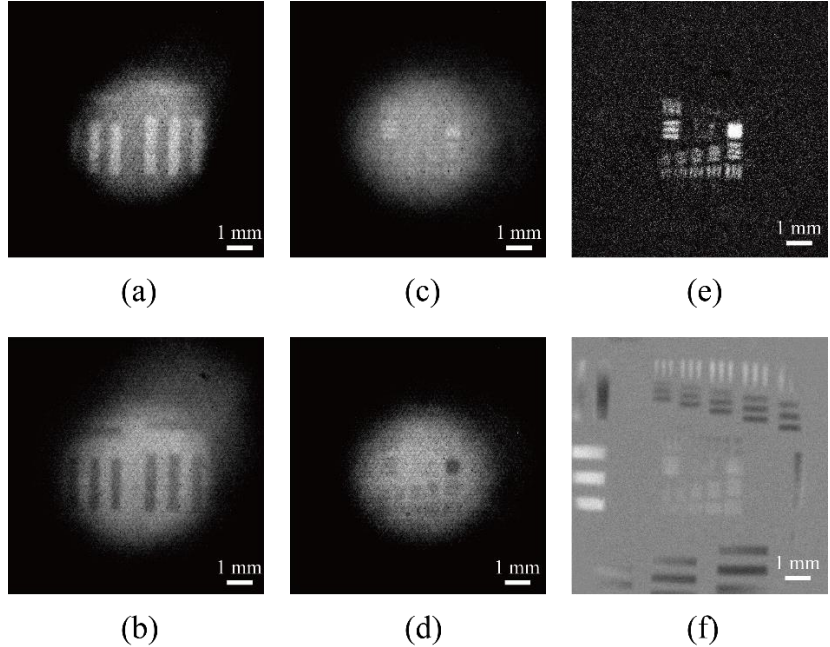

Fig. S2 Imaging results from the QNI. (a) and (b) constructive and destructive images of larger test patterns; (c), (d), and (e) constructive, destructive, and difference images of smaller test patterns; (f) interference image without band-pass filter used in the experiment.

## 3. Angular-spectrum-dependent (ASD) interference in the QNI

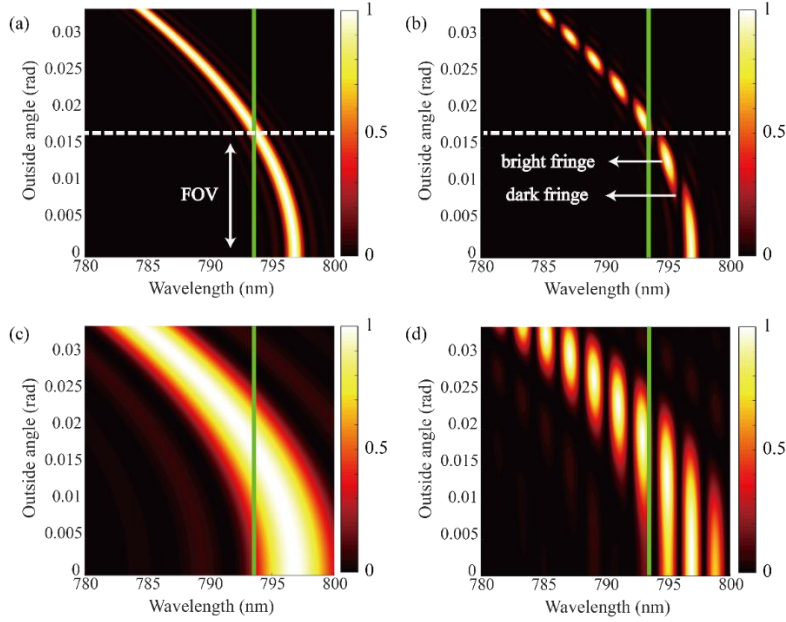

Fig. S3 (a) and (c) Frequency-angular spectrum $P(\omega_s,\theta_s)$, where the horizontal axis has been transformed into wavelength; (b) and (d) Integrand function $\left[1+\cos(2k_i d+\phi_1)\right]P(\omega_s,\vec{\kappa_s})$ where $d=160\,\mu m$. (a) and (b) Simulations based on a crystal with a length of 10 mm that was used in our experiment. (c) and (d) Simulations based on a crystal with a length of 2 mm.

In this section, we present supplementary results for the second case of OPD in section 4.3 in the main text. In Eq. (9), the photon counts on the detection plane are dependent on the distribution $P(\omega_s,\vec{\kappa_s})$. Here, considering the rotational symmetry around the direction of propagation of photons, one can only focus on the modulus of the transverse wave vector and the therefore two-variable function $P(\omega_s,\kappa_s)$. Then, the transverse wave vector $\kappa_s$ can be equivalently replaced by the emission angle $\theta_s$ outside the crystal based on the point-to-point mapping relation $\theta_s \approx \kappa_s/k_s \approx \kappa_s/k_{s0}$, where $k_{s0}$ is the modulus of wave vector of the center frequency. $P(\omega_s,\theta_s)$ is a frequency-angular spectrum of signal photons that shows the relation of frequency and outside angle [2]. Fig. S3 (a) shows the frequency-angular spectrum $P(\omega_s,\theta_s)$ in our experiment, which is calculated based on the non-collinear phase-matching condition [2, 3]. Substituting $P(\omega_s,\theta_s)$ into Eq. (9) and transforming the one-dimension integral result $\langle n(\theta_s)\rangle$ into two-dimension $\langle n(\sqrt{x^2+y^2}/f_2)\rangle$, one can obtain the interference patterns shown in Fig. 3 (a); an integrand function is shown in Fig. S3 (b) and one can directly know the position of bright and dark fringes in Fig. S4 (a) from the bright and dark region in Fig. S3 (b). The numbers on the top of Fig. 3 are arm differences corresponding to the patterns below them. We have demonstrated the path changes before lens L1 can cause equal-inclination interference in the main text, whereas the path changes after lens L1 just cause phase changes in far-field, therefore the equal-inclination interference does not occur here and the fringes are not caused by equal-inclination interference; they are caused by the combination of the interference patterns of the frequency components on the parabolic-shaped frequency-angular spectrum

$P(\omega_s, \theta_s)$. The phase $\phi_2 = \dfrac{b\omega_{s0}^2}{f_2^2 c^3}\rho_s^2 \Delta l$ in Eq. (6) is corresponding to this interference phenomenon, which is called angular-spectrum-dependent (ASD) interference here; $b = \dfrac{\omega_{s0}(\omega_{s0}n_s + \omega_{i0}n_i)c^2}{2n_i n_s \omega_i^3 (\beta_s \omega_{s0} + n_s - \beta_i \omega_{i0} - n_i)}$ [3], where $\beta_s = \left(\dfrac{dn}{d\omega_s}\right)_{\omega_{s0}}, \beta_i = \left(\dfrac{dn}{d\omega_i}\right)_{\omega_{i0}}$. Using Eq. (6), one can also obtain simulation results that are similar to those in Fig. 4 (a).

In the experiment, we use an 800-10 nm band-pass filter to reduce the number of fringes, where the center transmission wavelength is 800 nm and the FWHM bandwidth is 10 nm. The green vertical line in Fig. S3 (a) shows the cut-off wavelength of the filter on the short wavelength side. The intersection point of the green line and the parabolic-shaped frequency-angular spectrum gives the angle of field of view. Fig. S4 (b) shows the simulation results of the ASD interference with an 800-10 nm filter used (the approximate transmission function of the filter in the simulation is a super-Gaussian function $\exp\left[-\left(\dfrac{\lambda_s - 0.8}{0.0056}\right)^6\right]$, the unit of wavelength is micron ). Regretfully, in our experiment, we did not record the fringes in the case that the 800-10 nm band-pass filter is removed. We only recorded experimental data in the case that the filter is used and the results are shown in Fig. S4 (c). The experimental results can roughly agree with the simulations, the main difference between the simulation and experiment results is from the uncertain initial phase in the experiment; in the simulation, we accurately set the phase difference to make the center fringe brightness maximum, however, in the experiment we adjust the PZT to control the arm difference but the positions of maximum lack accuracy. Another reason causing the slight difference between experimental and simulation results is the dispersion of signal and idler wavelength in the lens L1. Actually, the interference pattern is not a uniform spot (the first figure in Fig. S4 (a)) without the filter used when the arm difference is zero; in the experiment, a large fringe can be seen in the ICCD in this case. Although we did not record the data, we find that the raw data in ref. [4] are similar to those we observed. Also, in Fig. S2 (f), one can see the interference pattern in the background and these background fringes are the reason why constructive and destructive interferences occur periodically along the radial direction. We deduce that the background fringes are caused by dispersion of signal and idler wavelengths because in our other experiment[3], the background fringes with zero arm difference did not occur, where we built a similar Michelson interferometer but not a nonlinear interferometer to study ASD interference using only signal photons and idler photons were discarded.

The ASD interference should be prevented in some cases, e.g. the QNI is used in phase imaging. A filter can be used in this case by cutting off the frequency-angular spectrum vertically, just as we do; an aperture can also work by cutting off the frequency-angular spectrum horizontally. When a filter or an aperture is used, all the photons in the field of view are in phase, just as shown in Fig. S2 (a)~(e). In addition, ASD interference can only occur when a long crystal is used. Because the visibility of the fringes is dependent on the width of the angular-frequency spectrum that is determined by the length of the crystal, a short crystal indicates low visibility of the ASD interference and therefore can prevent the ASD interference and meanwhile increase the field of view[5]. The simulation results using a crystal with a length of 2 mm are presented in Fig. S3 (c) and (d); one cannot obtain the interference fringes by integrating the function in Fig. S3 (d). On the other hand, when the ASD interference effect can be removed, e.g. using a difference technique or using a spatial light modulator to compensate the angle-dependent phase difference, a long crystal can be used in the imaging applications and the band-pass filter is unnecessary, where the large emission angle of SPDC can be used to enhance the field of view, just as Fig. S2 (f) comparing with Fig. S2 (e).

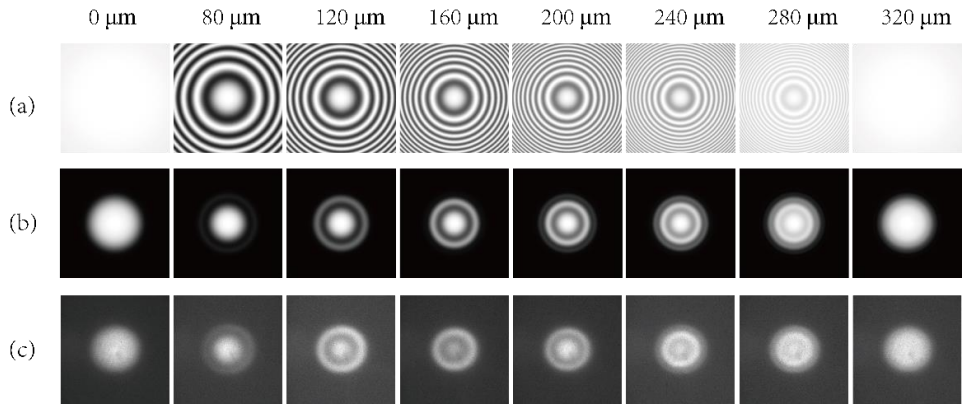

Fig. S4 (a) Simulation results for the ASD interference in our QNI. (b) Simulation results for the ASD interference with an 800-10 nm filter used. (c) Experimental results for the ASD interference with an 800-10 nm filter used. The numbers on the top are arm differences corresponding to the interference patterns below them.

Overall, the experimental results can roughly agree with the simulations, which indicates that the explanation (ASD interference) of the interference fringes is reliable. In ref. [3], we proved that the ASD interference has a much shorter equivalent wavelength, which means the number of fringes varies rapidly with increasing arm differences. This can also be observed in the QNIs. For example, in Fig. S4 (a), the fringes vary rapidly with arm difference within 300 μm, which is faster than those in usual equal-inclination interference, as well as the equal-inclination interference in a QNI, e. g. reported in the main text and ref. [6]. This phenomenon can be used to measure optical path changes within the coherence length of photons by recording variations of fringes, and the sensitivity is higher than equal-inclination interferometers because of the more sensitive dependence of fringes on displacement.